# Defining and Conceptualizing Actionable Insight: A Conceptual Framework for Decision-centric Analytics

**Shiang-Yen Tan**
School of Information Systems
Queensland University of Technology
Queensland, Australia
Email: s40.tan@qut.edu.au

**Taizan Chan**
School of Information Systems
Queensland University of Technology
Queensland, Australia
Email: t.chan@qut.edu.au

## Abstract

Despite actionable insight being widely recognized as the outcome of data analytics, there is a lack of a systematic and commonly-agreed definition for the term. More importantly, existing definitions are generally too abstract for informing the design of data analytics systems. This study proposes a definition of actionable insight as a multi-component concept comprising analytic insight, synergic insight, and prognostic insights. This definition is informed by a conceptual framework, which also can be used to systematically understand actionable insight, both at the concept-level and component-level. Each component is explained from the analytical, cognitive, and computational perspectives and relevant design considerations are suggested. We hope this study could be a rudimentary step toward the realization of decision-centric data analytics that can deliver the promised actionable insight.

## 1  Introduction

Data analytics involves the use of statistics, computational models, visualizations, and machine learning techniques to extract useful knowledge from large and complex datasets. The ideal of data analytics is to provide users with actionable insight, which enables the users to make decision that has pragmatic implications. For example, actionable insight from analysis of social media content and successes of prior marketing strategies enables the user to choose the right combination of marketing strategies to improve the company's image in a public relations crisis. In this notion, the purpose of data analytics is twofold. Firstly, data analytics help users to understand their data. Secondly, the higher-order purpose is to solve users' real-world problems through informed decision making.

The term "actionable insight" has been gaining much attention in both industry and academic in the past 5 years. In the industry, business executives, consultants, and software vendors have widely described actionable insight as the deliverable of data analytics software. More importantly, actionable insight is the key driver for businesses to invest in data analytics initiatives (Sawyer 2011). In academic, the term has been used in publications from different disciplines such as data mining, information visualization, business intelligence and analytics, and psychology.

Despite its popularity, there is a lack of a systematic and theory-driven definition of actionable insight. More importantly, existing definitions of actionable insight are generally too abstract to be used as guidance to inform the design and development of data analytics systems. Substantial data analytics systems to date are the results of advancement in data-driven and computational-oriented techniques. In contrast, systems that are designed to explicitly support users along the decision making process are scarce. Many senior executives and domain experts have commended that data analytics initiatives often failed to delivery practical value (Saraiya et al. 2005). One of biggest challenges faced by the practitioners is to intuitively apply the analytics results to solve a real-world problem (Houxing 2010). In other words, data analytics often failed to achieve its higher-order purpose which is to support the decision making. In accordance with these gaps, we contend there is need for a definition of actionable insight that can be used 1) to systematically understand actionable insight, and 2) to inform the design of decision-centric data analytics systems.

To address the needs, this study proposes a definition for actionable insight in the area of data analytics. This study conceptualizes actionable insight as a multi-component concept. Three components, namely 1) analytic insight, 2) synergic insight, and 3) prognostic insight, collectively, provide a holistic definition of actionable insight. These components are informed by a conceptual framework, namely HIVE. HIVE is grounded in theories from multiple disciplines to provide systematic understandings about actionable insight. At the component level, HIVE describes each component's informational, computational, and cognitive requirements. At the concept level, HIVE explains the interaction between the components and its overall characteristics.

This study hopes that the proposed definition can help both researchers and practitioners to establish a common and theory-driven understanding of actionable insight. More importantly, we hope the multi-component definition can be a starting ground for researchers to investigate component-specific gaps. For instance, user interactions, design principles, and assessment methods at each component need to be further explored. Although HIVE is currently in its rudimentary stage, our goal is to continuously refine the framework to be a native Information Systems (IS) theory that can be used to inform data analytics research.

## 2  Actionable Insight: A Multi-Component Definition

This study contends that in order to systematically define actionable insight, it is important to first understand actionable insight as a concept. One common theme in the somewhat fragmented definitions of actionable insight describes it as the knowledge which enables users to act upon meaningfully. In the context of data analytics, we conjecture that the actionable insight's capability to be acted upon meaningfully – or simply actionability, is to enable users to solve a problem in their domain through confident decision making. This study thus conceptualizes actionable insight as a cohesive set of reasoning artefacts that users gain along the data analytics process, with the purpose of solving a problem in the user domain. The literature review in this study was set to surround this central idea.

After reviewing theories and concepts from related disciplines which include complex problem solving, naturalistic decision making, sensemaking theory, and situation awareness, we conceptualize actionable insight as a multi-component concept which consists of three components, namely analytic insight, synergic insight, and prognostic insight. Such decomposition allows actionable insight to be examined in the way that is well aligned with the major processes and outcomes in the aforementioned theories. This enables complementarily explanations to be drawn from various disciplines to understand actionable insight from multiple perspectives. The resultant components of actionable insight result are:

a) **Analytic insight** – understanding and interpretation of individual analytical results.

b) **Synergic insight** – comprehension of the connections between the analytic insights and understanding of the problem situation by contextualizing the resultant comprehension in the user's objective, assumptions, and domain knowledge.

c) **Prognostic insight** – prediction of the problem situation's future states. Prognostic insight can also include the assessment of how the future states will change as the effects of different solutions, scenarios, and assumptions.

Based on the three insight components, this study proposes a definition of actionable insight as the following:

**Actionable Insight:** *A cohesive set of understandings about the problem situation based on prognostic insights derived from synergic understanding of analytical results which enables the user to make an informed decision to solve the problem.*

The set of understandings here refer to the reasoning artefacts the users gained at different stages of the data analytics process. Together, these progressive understandings enable the users to devise solution alternatives and decide on a solution that is best suit the user's objectives and scenario anticipated – thus, constitutes the "actionable" notion of the term. Another implication of the definition is that insight is more than merely technical analytic results. Insight is the result of information internalization process in which the analytic results are schematized according to certain mental models, contextualized in a specific situation, and reasoned using the user's analytical reasoning and heuristic.

## 3    Conceptual Framework of Actionable Insight

The definition of actionable insight introduce in the last section is informed by a conceptual framework, HIVE. The framework is developed to provide a systematic way to understand actionable insight, at both the concept-level and the component-level. Figure 1 illustrates the overview of HIVE framework and the relative position of insight components.

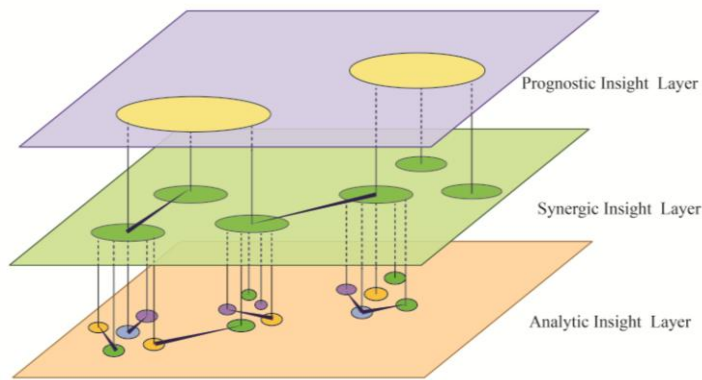

*Figure 1. Visual Representation of HIVE*

HIVE framework suggests that the insight components are hierarchical in nature. Components at the upper layers are built upon the components at the lower layers. Upcoming subsections describes each of these insight components in details, started with analytic insight and ended with prognostic insight.

### 3.1    Component1: Analytic insight

Each analytic insight is a relevant observation derived directly from one or more enquiries. Figure 2 shows the relation between analytic insight and enquiry. An enquiry can be in the form of database query, visualization, or mathematical computation that transforms data into information. An analytic insight may also derive from a series of sequential enquiries, which the latter enquiry is built on the former counterparts. For instance, data is first being clustered, and then the resultant clusters are used as inputs for association rule mining.

Analytic insight can be easily quantified and traced back to the enquiries which it being observed (Saraiya et al. 2005). Examples of analytic insight could be a set of association rules among other rules extracted, a significant pattern in a visual graph, or particular a relationship within a multi-regression model. Notice that the term "relevant observation" is used to imply only those observations which are relevant and meaningful in the user's current objective and context fit the notion of analytic insight.

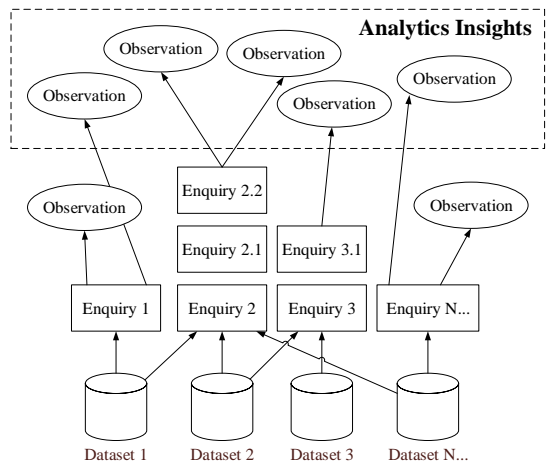

*Figure 2. From dataset to Analytic insight*

From the cognitive perspective, analytic insight is achieved when users successfully identify, perceives, and interprets relevant aspects of the data. Analytic insight involves the learnings about the status, attributes, and natures of the data subset. In the terms of sensemaking theory, users are engaging in foraging activities which include information searching filtering, extracting, and interpreting relevant information in order to gain analytic insight from data. In this data exploration stage, users are very inefficient in dealing with large and complex data due to the limited cognitive capability. Additionally, users tend to reduce the amount of environmental scanning (Chen and Lee 2003) and often explore the data based on hunches (David and Michelle 2009) when they deal with complex datasets. As the result, the data exploration often is a very time-consuming task that does not guarantee significant and meaningful observations. This challenge is particularly manifested in an ill-structured problem. At the early stage of such analysis, users often do not know what information to explore and everything could seem to be relevant to the analysis due to complex interrelations in the problem. Therefore, there is a need to support users in identifying variables that are relevant to current analysis context at the exploration stage. An example is to allow users to explore the data space in semantically meaningful structures. This allows them to 1) be aware of the interrelationships between the variables of interests and 2) discover other semantically related variables which otherwise remain hidden. The semantical structure can be graph-based knowledge map that produced from domain's ontology, results of machine learning, results of data mining, or the combination of these techniques.

**Design consideration:** Support users in identifying variables that are relevant to the current analysis context in order to improve the effectiveness of data exploration.

Among the other counterparts, analytic insight is a low-level insight. It involves highly objective information that require little judgemental heuristic. Each analytic insight has a narrow scope, as it focuses on a very specific piece of finding in the entire problem situation. Analytic insight reassembly the notion of using a high-power telescope to zoom at part of a city. It allows you to see extreme details of a building, but it's difficult to get an overview of the city layout and to find your way to the building. Alone, analytic insights are unlikely to carry direct implication for decision making or problem solving. In other words, the practical value of analytic insight is generally low. Nevertheless, most of the data analytics systems such as data mining, data visualization, and statistical analysis cease their supports beyond this point. In other words, these systems focus on fulfilling the immediate objective of data analytics which is to help the users to understand their data. However, the higher-order purpose to enable users to make informed decision is often not explicitly supported. Analytics activities beyond this point are often done in the user's mind, by paper-and-pen, or in other non-analytics systems such as mind-mapping tools. This also implies that the data analytics systems focus on low-level task that do not map well to the true needs of users, leaving the users contend with the gulf between low-level analytic insight and the big picture of the overall data analytics task.

## 3.2　Component 2: Synergic Insight

Synergic insight is comprehension about the entire problem situation based on the synthesis of individual analytic insights. Figure 3 illustrates the relations between analytic insights and synergic insight, intermediated by the chain of arguments (surrounded by the dotted rectangle in the middle). The bottom part of Figure 2 shows that an argument can be formed based on the synthesis of analytic insight(s) and soft evidence. Every argument focuses specifically on one topic, and draws one or more relevant analytic insight as the supporting evidence. An example of argument is "the decrease of competitive advantage in accounting software market". This argument can be supported by several analytic insights such as consistent drop in market shares (observed from data visualization) and the increase of unsatisfied customers (observed from text mining and sentimental analysis).

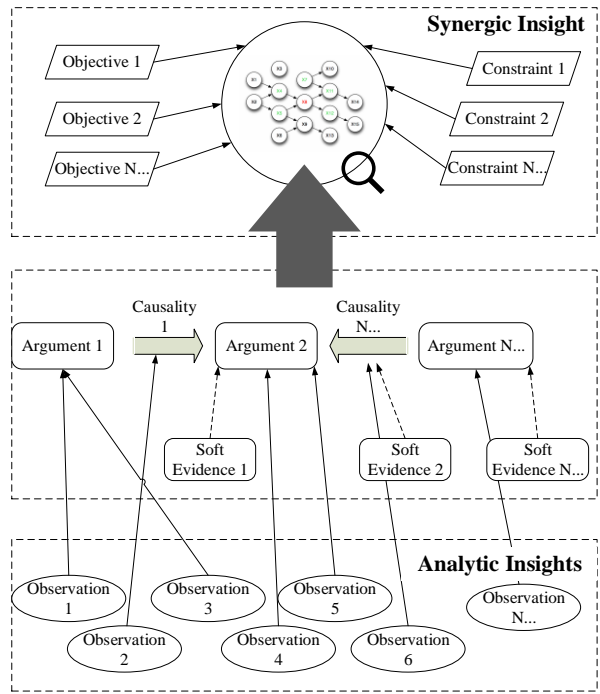

*Figure 3. From Analytic insights to Synergic Insight*

Besides the data-backed analytic insights that are used as the hard evidence, soft evidence can also be included to jointly support an argument. Soft evidences are various knowledge from users that are not captured in the datasets, such as domain knowledge (both explicit and tacit knowledge), judgemental heuristics, or logical reasoning. It is common that soft evidence is the glue that put the individual analytic insights into a cohesive argument (David and Michelle 2009). For example, the argument "decrease of competitive advantage in accounting software market" is also partially supported by a soft evidence, in which the board's meeting has led to the conclusion that the recent change in taxation policy will have negative impact on the competitive advantage the company possessed. Soft evidence can also come from the user's beliefs and confidence level toward the analytic insights. Notice that it is possible that an argument is formed merely based on soft evidence. This is especially true in real-world problem solving where certain arguments are vital, but yet the data to support them are impossible to obtain.

Causality is a specific type of argument that connects two or more arguments together to develop a chain of arguments. An example of chian of argument is "decrease in competitive advantage in personal accounting software market (an argument)" will negatively affect the "reputation of professional account software market (an argument)" through "reducing reputation of related products (a causality, which itself is also supported by one or more analytics insights or soft evidences). Subsequently, multiple chains of argument are schematized into a situation model to help users to comprehend the entire problem situation.

The schematization involves 1) organizing the chains of arguments by classification and rational representation and 2) contextualizing the chains of arguments in user's objectives and constraints. Studies in information systems have shown that explicit inclusion of objectives and constraints is vital for complex problem solving (Thomas and Cook 2005). The resultant situation model is an abstract system that represents the real-world situation which the users intend to solve.

From the cognitive perspective, synergic insight is achieved when users 1) understand the connections between the individual analytic insights and 2) comprehend the situation model as a whole. The situation model as a whole is larger than the sum of its analytic insights because the model unifies the overall connections to allow user to understand the dynamicity of the situation. An empirical study has affirmed the importance of synthesizing individual analytic insight in an analysis process. Their study shows over 80% of analysts reported that they synthesized two or two separate findings to form a joint conclusion. Situation model in this study can be seen as a part of the user's mental model that pertinent to the problem situation. Accurate mental model is one of the important factors to better strategic decision in business and complex problem solving (Albers 1999; Gary and Wood 2011). Similarly, situation model is the enabler of hypothesis generation, solution seeking, and other high-level cognitive activities that lead to effective problem solving (Ribarsky et al. 2009).

However, the construction of situation model is a daunting task because the problem often consists of conflicting objectives and complex connections. Research have also found that users often explore multiple situation models simultaneously (Gotz et al. 2006). Therefore, external aids for modelling the situation model is important. It helps to alleviate memory span issue and thus allows the user to focus on actual reasoning. The external representation of situation model provides a scaffolding for keeping all the chains of arguments, soft evidence, objective, and constraints in an integrated manner to be easily accessed by the user for reasoning.

**Design consideration:** Support the users in the construction, manipulation, storage, and representation situation models.

In terms of computational support, the importance of supporting users in synthesizing analytics results has gained increasingly awareness in data analytics, especially among the visual analytics community (Meso et al. 2002; Wright et al. 2006). A limited number of systems has started to support information synthesis by addressing the memory span limits of human users. However, most of the supports are not more than a canvas that requires users to manually organize and connect the individual analytic insights. The major drawback of the method is leaving the reasoning task to human user, which can be inefficient and subject to biases. There are many machine-aided reasoning techniques such as fuzzy cognitive maps, dynamic systems, and Bayesian belief network can be used to synthesize information and construct a structured model for reasoning purpose. To readily take the advantage of these techniques, the situation model and its components (e.g. chain of arguments, soft evidence, and annotation) must exist in the computer processable forms.

**Design consideration**: Support the users in constructing computation-friendly situation model, which can be further processed by computer-aided reasoning techniques.

In term of actionability, synergic insight does not readily warrant a decision. At this stage, users understand the dynamicity of the entire situation model based on the collective interactions among the elements. However, they do not know what would be the possible outcomes of the situation model in react to different solution alternatives or scenarios. Synergic insight is the vital foundation for users to design their solutions, and prepare for the most of important stage of problem solving which will be discussed in next section.

### 3.3  Component 3: Prognostic Insight

Prognostic insight is the predictions about the impacts of possible actions on the future state(s) of the problem situation. Prognostic insight is achieved when user is able to 1) generate hypotheses and subsequently 2) predict how the problem situation reacts to the hypotheses under different scenarios. Prognostic insight allows user to know future states of the problem situation. This enables users to know the ranges of possible outcomes and aware of the risk involved.

Figure 4 shows the how prognostic insight is derived from synergic insights. The synergic insight (i.e. situation model) is used to develop one or more hypotheses. There are two types of hypotheses can be developed. First type of hypothesis attempts to identify the most likely explanation to the situation without intervention. Different possible explanations are acting as competing hypotheses. Second type of hypothesis involves intervention, in which different possible actions are devised by users and then incorporated into the situation model. An example of an action could be "provide free web seminars to users in the trial period". The intention is to examine what are the effects of these actions on the other arguments in the situation model and eventually their impacts on the user's objectives. Multiple actions can be grouped into one hypothesis. In this case, different sets of action are acting as the competitive hypotheses.

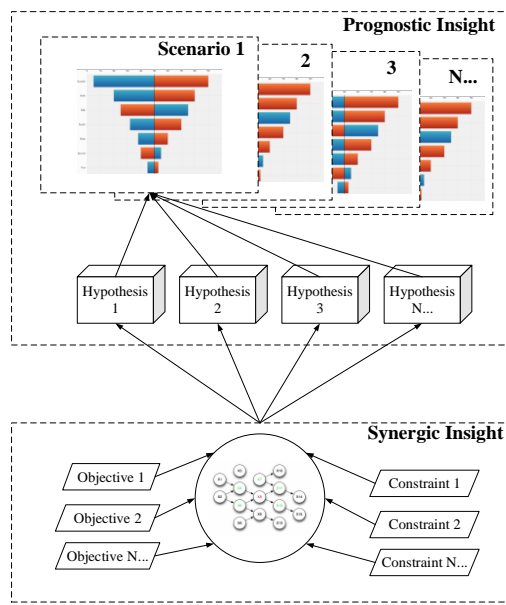

*Figure 4. From Synergic Insight to Prognostic Insight*

To gain prognostic insight, the hypotheses are then assessed under different variations of the situation models, namely scenarios. The variations can be at the argument level (i.e. changes of the argument's parameter), model level (i.e. inclusion or exclusion of certain arguments and causalities), or the problem level (i.e. changes of objectives). The hypotheses are simulated under different scenarios to understand the how the scenarios affect the hypotheses. The complexity of the prognostic insight can grow exponentially and beyond manageable as variation of scenarios increase. Additionally, complex problems often have no true or false solution and multiple solutions may exist. Research has shown that users can be benefited in such complex problem situations by engaging in gaming-like processes to clarify the nature of the problem (Mirel 2004). In other words, this requires a blend of experimental and analytics techniques to deal with uncertainty and dynamic natures of the problem situation.

**Design consideration:** Support the modelling, management (i.e. save and retrieve), analysis, and comparison of multiple hypotheses, scenarios, and situation models.

From the cognitive perspective, users want to "run" their situation model to understanding the dynamicity of model by performing variety of "what if" analyses. As the outcome, it change the decision makers' understanding about how the problem situation works and compel them to reorganize their mental model (Weick 1995). These activities involve cognitive taxing activities such as mental simulation and scenarios building. This is a daunting task because problem situation often characterizes high level of uncertainty and involves large number of interrelated factors. Added to the complication, these factors are time-sensitive, have non-linear relationships that render the problem situation changes in a complex way. Apart from that, probabilistic analysis becomes inevitable in such problem situation with high level of uncertainty. Nevertheless, human is proven to be weak at probabilistic analysis without external mathematical aids. In addition to manage, holding, and visualize the situation, advanced analytics techniques such as machine learning and computer-aided reasoning can be incorporated to augment the reasoning process. Such structured reasoning encourages rigorous and logical processing which able to enhance the validity of the reasoning outcomes and reduce cognitive pitfalls.

**Design consideration:** Support structured reasoning with the aids of advanced analytics techniques.

Prognostic insight has the highest extent of actionability as it provides users with the knowledge necessary to decide on the most favourable course of actions to meet their objective. By simulating the hypotheses under different scenarios, users are able to assess the range of potential outcomes and the associated risk. Prognostic insight also enables the users to adopt anticipatory strategy to the problem situation with a faster cycle of insight creation, which is critical in complex and dynamic environment typical of modern organizations (Chen and Lee 2003).

# 4　Overall Characteristics and Interactions between Components

## 4.1　Characteristics of the Insight Components

The three insight components characterize different extents of abstractions, content granularity, objectivity, human reasoning, and domain value. Figure 5 summarizes the characteristics of three components.

| Insights＼Aspects | Analytical Insight | Synergic Insight | Prognostic Insight |
|---|---|---|---|
| Abstraction level | | | |
| Scope | | | |
| Granularity | | | |
| Objectivity | | | |
| Human Reasoning | | | |
| Domain Value | | | |

*Figure 5. Characteristics of the Three Insight Components*

The level of abstraction and content granularity are reversely associated. The higher the abstraction level, the lower the content granularity. Prognostic insight at the top layer characterizes highest level of abstraction and widest scope, while has the lowest level of content granularity as indicated by the first three rows in Figure 5. A prognostic insight usually concerns about interactions of various high-level factors in the entire business ecology, such as consumer purchasing trend and changes in an international trading policy. At this level, the granularity is relatively low as the detail data such as sales volume of different product lines grouped by regions is no longer an input for the reasoning at this level.

Objectivity of the insights is high at the lower layer and low in the higher layer. For instance, analytic insights at the lowest layer are mainly concerning quantitative and objective information such as technical indices produced by analysis tests. Subjectivity of the insight increases toward the higher end as more qualitative is being integrated. Subjective information such as domain knowledge, contextual information, and judgmental heuristics become more important and prevalent toward the higher layers.

The higher the insight located in the framework, the more important the human reasoning in deriving that insight. Information at higher-level often requires complex processing such as synthesizing information, extracting semantic meanings, and generating hypotheses. Such qualitative reasoning is the weakness of conventional computational methods. Relatively, human reasoning and judgmental heuristic allow users to perform effectively in deriving higher-level insights such as synergic and prognostic insights. In other words, the workloads on human reasoning increase toward the higher end of the insight layers.

Domain value tends to increase toward the higher layers. In other words, high-level insights such prognostic insight is relative readily to be translated into practical decisions or actions. Insights on the higher layers tend incorporates more domain or context specific information into the analysis, such as user's current objective, constraints, and solution alternatives.

## 4.2 Interactions between the components

There are interdependent relationships between the three components of actionable insight. The relationships can be described as 1) non-sequential and 2) iterative.

Although the three components are hierarchically related, the processes to produce them are not necessarily sequential. Users can gain the insight components in any order. In a deductive order, users can work from prognostic insight (i.e. preliminary hypothesis about how the situation will react to the solutions the users derived based on their experience) to analytic insight (i.e. to find evidence to support his conjecture). In an inductive order, users might work with analytic insights to build up his understanding of his problem situation, identify the potential problem, and devise potential solutions.

Very commonly, users work in an iterative pattern. For instance, the users first quickly build a rough situation model and execute the model to gain initial prognostic insight. Then, based on the feedbacks, they gradually refine the situation model by exploring new analytic insights. After they obtained a satisfactory situation model, they will attempt to gain prognostic insight from it again. To effective support a data analytics task in a real-world situation, the systems should support the free-flowing thoughts of the users, rather than constrains the flow-of-thoughts within certain rigid process.

**Design consideration**: to support user's natural analytics behaviours. Specifically, to provide flexible analytics environment that supports the natural flow-of-thoughts of the users and also the fluid switch from working on one insight component to another with minimal cost.

## 4.3 Processes and Supports for the Insight Components

This subsection presents 1) the processes in which the insight components are resulted and 2) existing system supports to achieve the insight components. Figure 6 shows three insights components and how well they are supported by existing data analytics software and research. Green colour indicates the well-supported areas, whereas the grey colour indicates the areas that are under-supported.

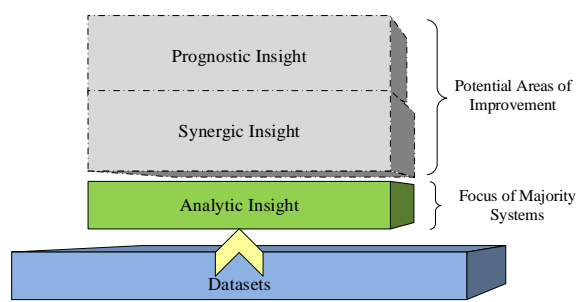

*Figure 6. Insight Components and Existing Supports*

Starting from the bottom, analytic insight is the outcome of data exploration stage. In data exploration stage, users transform the data in order to answer "what", "when", "how much", and possibly "why" questions. Software support and research at the analytic insight layer are comprehensive and matured. However, most of the works have been focused on the technical-driven advancement. There is a need for user-driven exploratory techniques which enable effective data exploration. One way to achieve this is to provide relevant clues and knowledge-assisted guidance that lead to efficient and accurate data exploration (Shrinivasan 2010; Xiaoyu et al. 2011).

Synergic insight is the outcome of information synthesis stage. The main activities in this stage includes abstracting, organizing, assigning semantic meanings to analytics result, and produce new knowledge based on joint findings (Robinson 2009). The purpose of information synthesis is to connect those fragmented analytical results into big picture that reflect the overall analytical task. The lacking of supports for information synthesis is a well-recognized gap in the field of analytics. Conceptual research at the synergic insight level has been long existed. However, there is a real need for developing techniques that can realize the conceptual designs. The efforts should go beyond the conventional "evidence marshalling" or "evidence shoebox" which relies on human users for connecting, organizing, and reasoning (Thomas and Cook 2005). Automation of the processes in this phase using techniques such as case-based reasoning and artificial intelligence could improve the effectiveness of the model construction and reduce human-induced biases.

Prognostic insight is the outcome of structured reasoning stage. Structured reasoning is central to the analytical task of applying human judgment to reach conclusion or devise solution (Ribarsky et al. 2009; Shrinivasan and Wijk 2008). Prognostic insight is the layer which requires most research works and software supports. Most of the prediction and simulation techniques are meant for quantitative data and poor in dealing with uncertain, dynamics, and subjective data. This study contends that various fuzzy and hybrid methods, such as fuzzy cognitive maps, Bayesian network, and dynamic systems can be incorporated into data analytics for supporting this high-level reasoning task.

## 5 Conclusion

Although this study is still in its rudimental phase, we believe the main contributions are residing in the proposed definition and the conceptual framework of actionable insight. We hope the definition in this study can establish a common expectation for the term "actionable insight" and thus conceptually lay out what is needed to be achieved by data analytics systems in order to enabling actionable insight. Additionally, we wish to bring to the attention that there are research opportunities at the synergic and prognostic insights levels. The design requirements, design principles, functionalities, and evaluation methods at these levels are yet to be fully studied. As for the HIVE framework, we hope that researchers can use it as a basis for deriving design principles and system functionalities to realize decision-centric data analytics systems. Moreover, it can be used a conceptual architecture for integrating existing analytics techniques into a seamless analytics process model.

This study does not claim the proposed definition as the only definition of actionable insight. However, we believe the multi-components definition is one of the perspectives which data analytics researchers can adopt to enrich existing understandings about actionable insight. Additionally, HIVE framework is not meant to inclusively classify all types of insight and does not deny the existence of insights besides what have been described in the framework. For example, navigational insights which enable users to know what are the next analysis steps to adopt are not within the scope of the framework.

The authors are using the HIVE framework as the guiding principles for developing a decision-centric data analytics system. Conjunction with the HIVE structure, the system embraces the underlying principles to allow users to have a one-stop solution for achieving analytic insight, synergic insight, and prognostic insight in single integrated platform.

The study hopes to further refine HIVE framework into a native IS theory that can be used by IS researchers as a kernel theory to inform the design and development of data analytics systems. Authors would highly welcome other researchers to expand, correct, and improve the proposed HIVE framework to be a better conceptual framework for the data analytics community.

# 6　References

## Copyright